\documentclass{article}
\usepackage{subcaption}

\usepackage{arxiv}

\usepackage[utf8]{inputenc} % allow utf-8 input
\usepackage[T1]{fontenc}    % use 8-bit T1 fonts
\usepackage{hyperref}       % hyperlinks
\usepackage{url}            % simple URL typesetting
\usepackage{booktabs}       % professional-quality tables
\usepackage{amsfonts}       % blackboard math symbols
\usepackage{nicefrac}       % compact symbols for 1/2, etc.
\usepackage{microtype}      % microtypography
\usepackage{lipsum}
\usepackage{graphicx}
\graphicspath{ {./images/} }

\title{Alias4SBML: A Python Package for Generating Alias Nodes in SBML Models}

\author{
Adel Heydarabadipour \\
  Department of Bioengineering\\
  University of Washington\\
  Seattle, WA 98195-5061 \\
  \texttt{adelhp@uw.edu} \\
   \And
Herbert M Sauro \\
  Department of Bioengineering\\
  University of Washington\\
  Seattle, WA 98195-5061 \\
  \texttt{hsauro@uw.edu} \\
}

\begin{document}
\maketitle
\begin{abstract}
Interpreting biological networks becomes challenging when molecular components, such as genes or proteins, participate in numerous interactions, resulting in densely connected regions and overlapping interactions that obscure functional relationships and biological insights. To address this, we introduce \textbf{Alias4SBML}, a Python package that enhances SBML model visualizations by generating alias nodes—duplicate representations of highly connected molecular components—to redistribute interactions and reduce visual congestion. Applying Alias4SBML to the SBML models, including one with 59 species and 41 reactions and another with 701 species and 505 reactions, demonstrated significant improvements in readability, with edge length reductions of up to 50.88\%. Our approach preserves the structural integrity of the network while facilitating clearer interpretation of complex biological systems, offering a flexible and scalable solution for visualizing biological models more efficiently.
\end{abstract}


\section{Introduction}
Visualization is an essential tool in scientific communication, transforming complex biological data into clear, interpretable representations that aid in understanding and discovery \cite{hoksza2020closing}. Two-dimensional pathway maps and network visualizations are particularly valuable, providing researchers with a structured view of intricate biological systems \cite{king2015escher}. However, one major challenge with these visualizations arises when molecular components, such as genes or proteins, participate in numerous interactions. This results in densely connected regions that obscure functional relationships and biological insights. Specifically, in large-scale models, the overlap of molecular components and their interactions often leads to cluttered and tangled visual representations, which complicate the interpretation of the model.

To address this challenge, we introduce \textbf{Alias4SBML}, a Python package that generates alias nodes—duplicate representations of highly connected molecular components. These alias nodes redistribute interactions more evenly across biological networks formatted in SBML (Systems Biology Markup Language) \cite{hucka2003systems}. By alleviating visual congestion, alias nodes preserve the clarity of functional relationships and enhance the overall interpretability of biological networks. The module integrates seamlessly with SBMLDiagrams \cite{xu2023sbmldiagrams}, a model visualization tool, to generate visualization data for SBML models that lack predefined layouts. This integration improves the clarity of network visualizations. Additionally, Alias4SBML supports configurable alias node generation, allowing users to specify which molecular components should have alias nodes and define connectivity thresholds, ensuring adaptability across a wide range of biological models.

\section{Methodology}

We implemented a structured workflow to enhance the visualization of biological networks through the generation of alias nodes. This process involves importing SBML models, generating visualization data when necessary, configuring aliasing parameters, generating alias nodes, redistributing interactions, and exporting the model with its enhanced visualization data. The following subsections outline each step of this workflow, detailing how the tool processes SBML models and applies alias node generation to improve visualization clarity.

\subsection{Loading SBML Model} 
SBML models can be imported into the module from either file input or directly from SBML or Antimony \cite{smith2009antimony} strings. To convert Antimony to SBML models, we utilize Tellurium \cite{choi2018tellurium}. This flexibility allows users to work with models in different formats, accommodating various modeling workflows.

\subsection{Adding Visualization Data}
If an SBML model does not include predefined visualization data, Alias4SBML integrates with the SBMLDiagrams package to automatically generate Layout \cite{gauges2015systems} and Render \cite{bergmann2018sbml} data—SBML extensions that store its visualization data. This step ensures that the model contains the necessary visualization data, which is a prerequisite for alias node generation.

\subsection{Settings Configuration}
The user has the flexibility to configure alias node generation for either all species in the model or for specific components of interest. This selective approach enables users to focus aliasing on the most complex regions of the network, preventing unnecessary clutter in the visualization. 

Additionally, the module allows users to define a connectivity threshold, specifying the maximum number of interactions a node can have before an alias node is generated. This feature provides users with granular control over the aliasing process according to their specific needs.

\subsection{Alias Node Generation}
Once the visualization data is prepared and aliasing parameters are set, Alias4SBML identifies species that meet the predefined aliasing criteria. For each selected species, the module generates alias nodes. The module ensures that alias nodes are appropriately positioned within the visualization layout to maintain clarity in network interpretation. The interactions (connected edges) associated with the original species are then redistributed between the original and alias nodes, effectively reducing edge density in highly connected regions. This approach mitigates visual clutter while preserving the structural and functional integrity of the model. 

\subsection{Saving the Model}
The final network representation, enhanced with alias nodes, can be exported in SBML format with its visualization data stored in the Layout and Render extensions.

\section{Results and Discussion}

To evaluate the efficacy of alias node generation in enhancing network visualization, we applied our approach to several SBML models and compared their representations before and after the introduction of alias nodes.

\begin{figure}[t]
    \centering
    \begin{subfigure}[b]{0.475\textwidth}
        \centering
        \includegraphics[width=\textwidth]{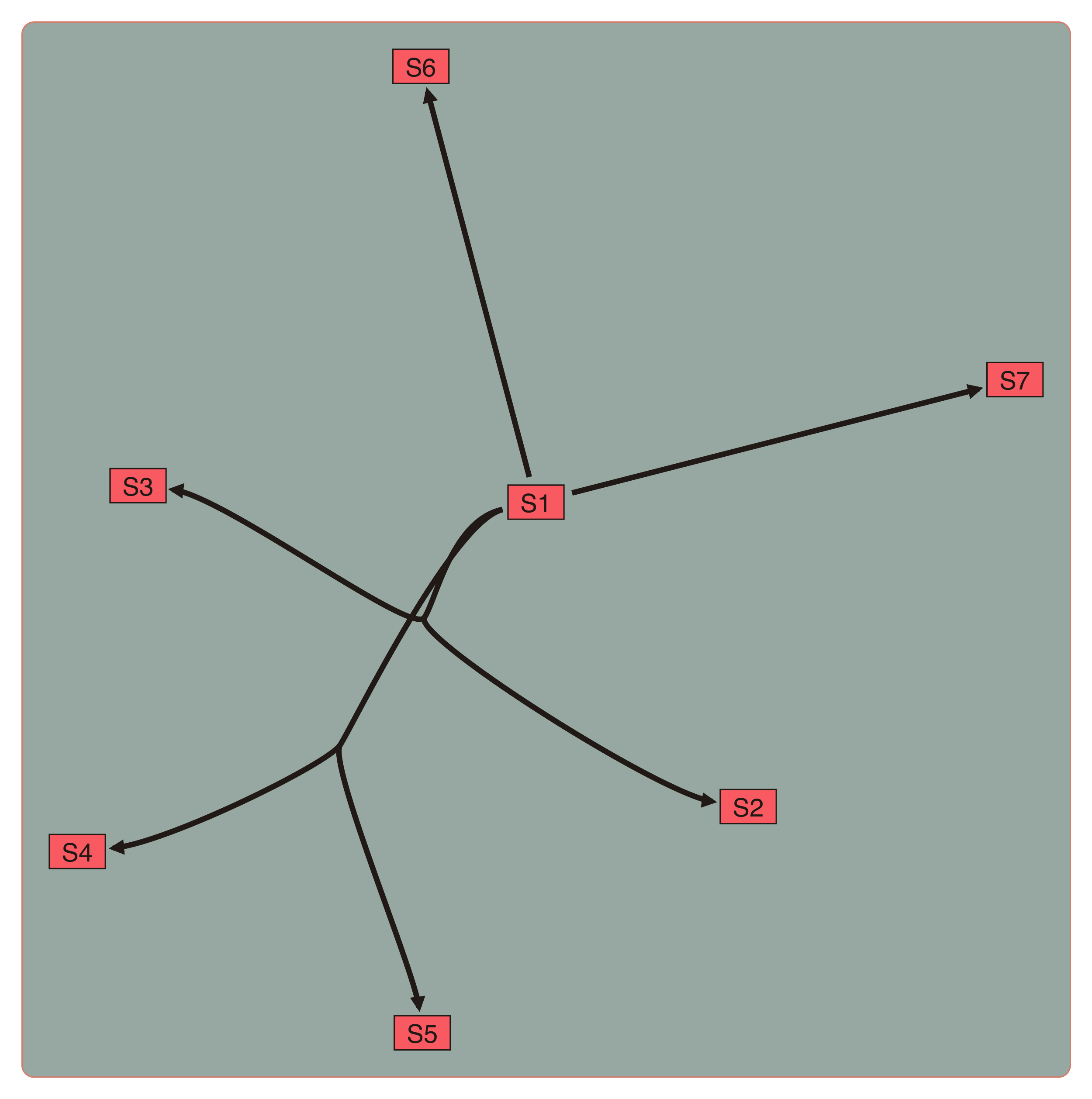}
    \end{subfigure}
    \hspace{0.02\textwidth}
    \begin{subfigure}[b]{0.475\textwidth}
        \centering
        \includegraphics[width=\textwidth]{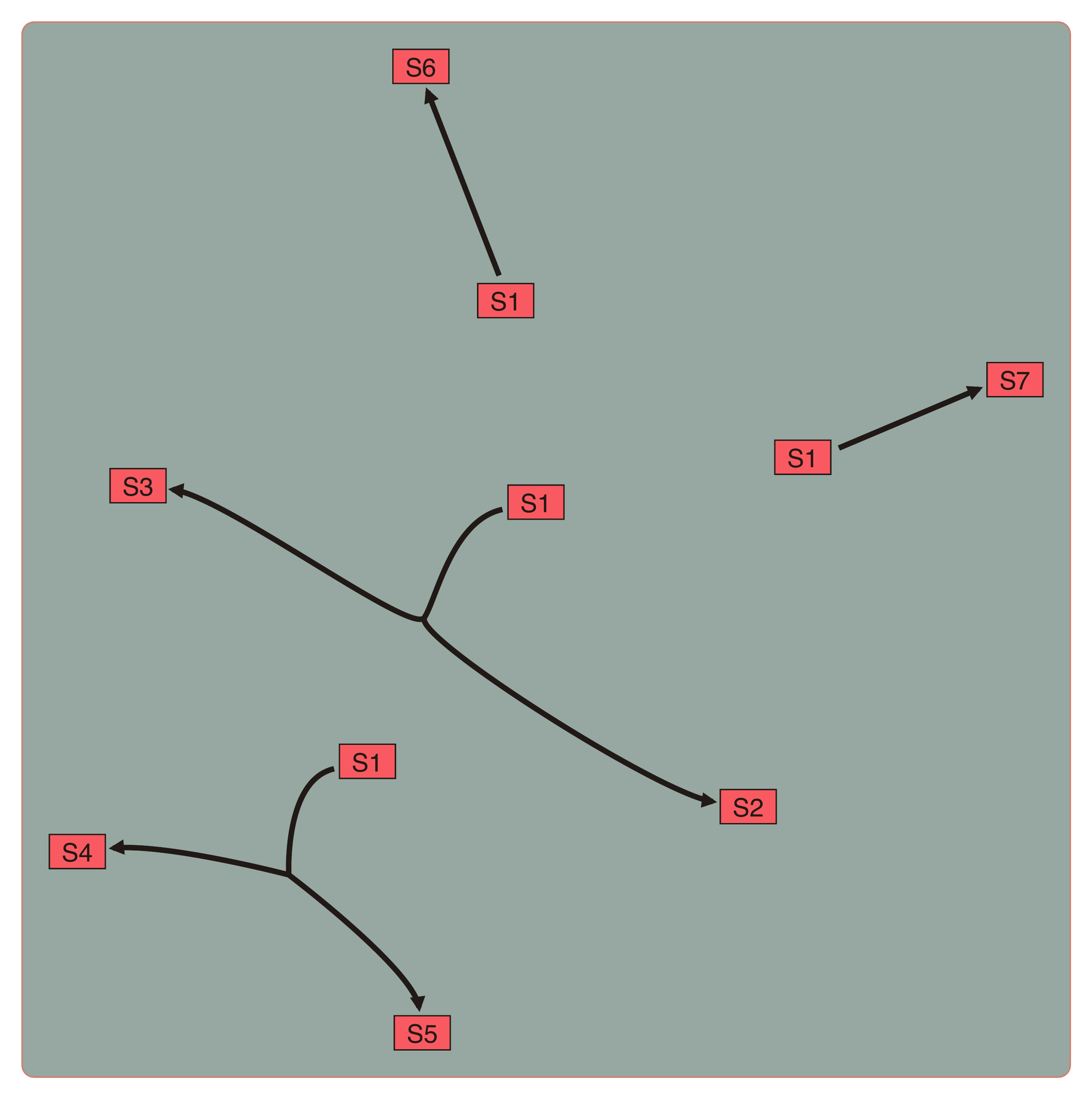}
    \end{subfigure}
    \caption{Visualization of a network containing a highly connected species (\texttt{S1}) before and after alias node generation. \textbf{Left:} Initial visualization showing \texttt{S1} involved in multiple interactions, causing edge overlap and reduced clarity. \textbf{Right:} After aliasing, interactions are redistributed, alleviating visual congestion and improving network readability.}
    \label{fig:before_after}
\end{figure}

In the first example (Figure~\ref{fig:before_after}), we examined a model containing a highly connected species, \texttt{S1}, which participates in multiple interactions. For this example, alias node generation was configured to apply only to species \texttt{S1}, with a connectivity threshold set to 1, meaning that alias nodes were introduced only when \texttt{S1} had more than one interaction. In the initial visualization, the high connectivity of \texttt{S1} resulted in significant edge overlap, obscuring individual interactions and complicating the interpretation of the overall network structure. Following the introduction of alias nodes, interactions were redistributed, reducing edge congestion and preventing excessive bundling around \texttt{S1}. This modification substantially improved the visual clarity of reaction pathways and produced a more interpretable network representation.

\begin{figure}[t]
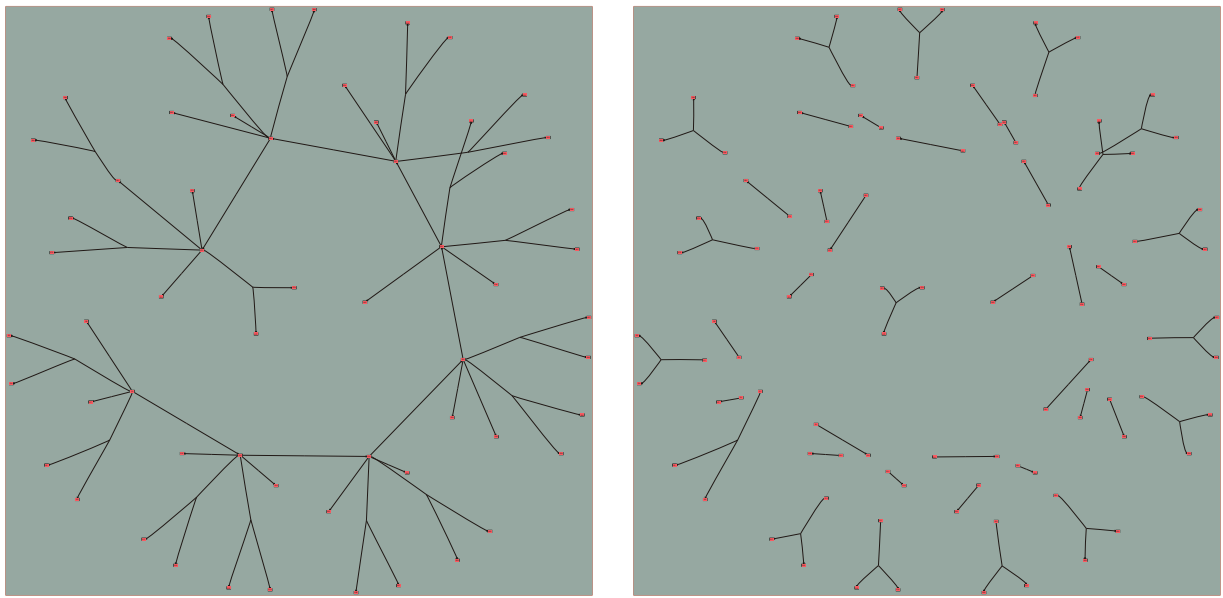

    \centering
    \begin{subfigure}[b]{0.475\textwidth}
        \centering
        \includegraphics[width=\textwidth]{large_before.pdf}
    \end{subfigure}
    \hspace{0.02\textwidth}
    \begin{subfigure}[b]{0.475\textwidth}
        \centering
        \includegraphics[width=\textwidth]{large_after.pdf}
    \end{subfigure}
    \caption{Visualization of an SBML model with 59 species and 41 reactions before and after alias node generation. \textbf{Left:} Initial visualization showing high edge overlap due to the connectivity of multiple species. \textbf{Right:} After aliasing, interactions are redistributed, resulting in a 47.3\% reduction in edge length and improved network clarity and readability.}
    \label{fig:before_after2}
\end{figure}

To further assess the scalability of our approach, we applied it to a more complex SBML model comprising 59 species and 41 reactions (Figure~\ref{fig:before_after2}). A comparison of the visualized network before and after alias node generation revealed a marked improvement in readability. We quantified this improvement by measuring the total length of all interaction paths in the network both before and after aliasing, and calculated the reduction percentage as follows:

\[
R = \frac{L_{OM} - L_{AM}}{L_{OM}} \times 100
\]

\textbf{where:}
\begin{itemize}
    \item \( R \) is the \textbf{Interaction Path Length Reduction Percentage},
    \item \( L_{OM} \) is the \textbf{Total Interaction Path Length in the Original Model} (i.e., the sum of all interaction path lengths in the unmodified SBML visualization), and
    \item \( L_{AM} \) is the \textbf{Total Interaction Path Length in the Aliased Model} (i.e., the sum of all interaction path lengths after alias node introduction).
\end{itemize}

For the SBML model with 59 species and 41 reactions, the \( R \) value was 47.3\%. To further examine the scalability of our module, we applied it to a larger model containing 701 species and 505 reactions, for which \( R \) was calculated to be 50.88\%. These results highlight the scalability of alias node generation, demonstrating its consistent ability to improve network readability, even as model complexity increases, by effectively redistributing interactions without disrupting the fundamental structural relationships among species.

The observed improvements suggest that alias node generation is an effective strategy for mitigating visual clutter in SBML model visualizations. By reducing edge overlap, alias nodes facilitate a clearer separation of reaction pathways, improving the overall interpretability of complex biological networks. Although aliasing effectively reduces visual clutter, it can simplify certain aspects of the network's structure. Therefore, fine-tuning parameters—such as adjusting connectivity thresholds and selectively choosing which species to alias—is essential to striking a balance between enhanced visual clarity and the preservation of critical biological information.

Overall, these findings demonstrate that \textbf{Alias4SBML} provides a robust tool for enhancing the visualization of biological networks. The significant improvements in network readability, as evidenced by the reduction in total interaction path length, highlight the potential of alias node generation as a valuable technique for biological model visualization. Future work will focus on refining the aliasing process and developing adaptive methods to dynamically adjust parameters based on local network density, further broadening the applicability of our approach.

\section{Conclusion}

In this study, we introduced \textbf{Alias4SBML}, a Python package developed to enhance the visualization of biological networks by generating alias nodes in SBML models. Our approach effectively addresses the challenges posed by densely connected networks. The results demonstrate that alias node generation significantly reduces edge overlap and improves the separation of reaction pathways, thereby enhancing overall network readability while preserving the integrity of essential structural relationships. The scalability of our approach was demonstrated across models of varying complexity, with consistent improvements observed in both small- and large-scale networks. These findings highlight the potential of Alias4SBML as a valuable tool for researchers working on biological models, offering a means to facilitate clearer interpretation and communication of complex biological interactions. Future developments will focus on refining the aliasing process further and incorporating additional customization options to meet the diverse needs of the scientific community.

\section{Software Availability and Usage}
The source code for \textbf{Alias4SBML} is openly available on GitHub at \url{https://github.com/adelhpour/Alias4SBML}. The Python package can be easily installed via \texttt{pip} using the following command:
\begin{verbatim}
pip install alias4sbml
\end{verbatim}

A simple example demonstrating the use of Alias4SBML is provided below:
\begin{verbatim}
import alias4sbml

model = '''
J0: S1 -> S2;
J1: S1 -> S3;
J2: S1 -> S4 + S5;
'''

a4sbml = alias4sbml.load(model)
a4sbml.create_alias(species="S1", max_species_connections=2)
a4sbml.draw("aliased_model.png")
\end{verbatim}

This example illustrates how to load an SBML model, configure alias node generation for a specific species, and generate an enhanced visualization. All code used to generate the data and figures presented in this paper is available in the same GitHub repository.

\section{Acknowledgements}

This work was supported by the National Institutes of Health under award number U24EB028887. The content expressed here is solely the responsibility of the authors and does not necessarily represent the official views of the National Institutes of Health or the University of Washington.

\bibliographystyle{unsrt}  % You can choose a different style if needed
\bibliography{references}  % This will include your references from the references.bib file

\end{document}